\documentclass[12pt]{article}
\usepackage{graphicx}
\usepackage{amsmath}
\usepackage{multirow}
\usepackage{amssymb}
\usepackage{natbib}
\usepackage{moreverb}
\usepackage[T1]{fontenc}
\usepackage{tkz-graph} 
\usetikzlibrary{positioning, shapes}
\usepackage{gensymb}
\usepackage[utf8]{inputenc}
\usepackage{hyperref}
\usepackage{pdfpages}
\usepackage[figuresright]{rotating}
\usepackage{geometry}
\usepackage{bbm}
\usepackage{indentfirst}
\usepackage{tablefootnote}
\usepackage{amsthm}
\usepackage{setspace}
\renewcommand{\P}{{\mathbb{P}}}
\newcommand{\E}{{\mathbb{E}}}

\newcommand{\x}{{\mathbf{x}}}

\newcommand{\bof}{{\mathbf{f}}}
\newcommand{\bF}{{\mathbf{F}}}

\newcommand{\btau}{{\boldsymbol\tau}}

\newcommand{\brho}{{\boldsymbol\varrho}}
\newcommand{\bu}{{\mathbf{u}}}
\newcommand{\bb}{{\mathbf{b}}}

\newcommand{\Var}{{\mathbf{Var}}}

\newcommand{\z}{{\mathbf{z}}}
\newcommand{\Z}{{\mathbf{Z}}}

\newcommand{\bR}{{\mathbf{R}}}
\newcommand{\bZ}{{\mathbf{Z}}}

\newcommand{\cF}{{\mathcal{F}}}

\newcommand{\br}{\mathbf{r}}
\newcommand{\bq}{\mathbf{q}}

\title{\textbf{Sensitivity Analysis for Multiple Comparisons in Matched Observational Studies through Quadratically Constrained Linear Programming}}
\author{Colin B. Fogarty  \and Dylan S. Small\thanks{Colin B. Fogarty is Doctoral Candidate, Department of Statistics, The Wharton School, University of Pennsylvania, Philadelphia PA 19104 (e-mail: \texttt{cfogarty@wharton.upenn.edu}).  Dylan S. Small is Professor, Department of Statistics, The Wharton School, University of Pennsylvania, Philadelphia PA 19104.}}
\date{}

\begin{document}
\maketitle
\thispagestyle{empty}
%

\abstract{A sensitivity analysis in an observational study assesses the robustness of significant findings to unmeasured confounding. While sensitivity analyses in matched observational studies have been well addressed when there is a single outcome variable, accounting for multiple comparisons through the existing methods yields overly conservative results when there are multiple outcome variables of interest. This stems from the fact that unmeasured confounding cannot affect the probability of assignment to treatment differently depending on the outcome being analyzed. Existing methods allow this to occur by combining the results of individual sensitivity analyses to assess whether at least one hypothesis is significant, which in turn results in an overly pessimistic assessment of a study's sensitivity to unobserved biases. By solving a quadratically constrained linear program, we are able to perform a sensitivity analysis while enforcing that unmeasured confounding must have the same impact on the treatment assignment probabilities across outcomes for each individual in the study. We show that this allows for uniform improvements in the power of a sensitivity analysis not only for testing the overall null of no effect, but also for null hypotheses on \textit{specific} outcome variables while strongly controlling the familywise error rate.  We illustrate our method through an observational study on the effect of smoking on naphthalene exposure.}

\vspace{.15 in}
\newpage
\setcounter{page}{1}
\section{Introduction}
\subsection{Unmeasured Confounding with Multiple Outcomes}
Conclusions drawn from an observational study should be subjected to additional scrutiny due to their vulnerability to unmeasured confounding. Unlike with a randomized experiment, a covariate which has not been adjusted for in the primary analysis may very well drive the observed relationship, thus nullifying the study's original finding. This necessitates an additional step known as a \textit{sensitivity analysis} to assess the robustness of an observational study's conclusions. A sensitivity analysis seeks an answer to the following question: how extreme would hidden bias have to be in order for the conclusions of a study to be materially altered? A study whose findings could be overturned with a small amount of unmeasured confounding invites warranted skepticism, while a study's conclusions are bolstered if a large degree of unmeasured confounding is required.

A sensitivity analysis computes worst-case bounds on the desired inference at a given level of unmeasured confounding. In observational studies employing matching to adjust for overt biases, the corresponding sensitivity analysis has been well studied when there is a single outcome variable of interest; see Section 4 of \citet{obs} for a comprehensive overview. It is parameterized by a number $\Gamma \geq 1$ which controls the allowable departure from purely random assignment for individuals who are similar on the basis of their observed covariates: two individuals in the same matched set can, due to the presence of unmeasured confounding, differ in their odds of assignment to treatment by at most $\Gamma$. Higher values of $\Gamma$ thus allow for unmeasured confounding to more substantially bias the treatment assignment probabilities for individuals in the same matched set. As discussed in Section \ref{sec:sens}, the impact of unmeasured confounding can be encoded by a scalar latent variable, $u_{ij}$, which represents the aggregate impact of unmeasured confounding on the assignment probabilities for individual $j$ in matched set $i$.  Individuals in the same matched set with higher values for $u_{ij}$ have higher probabilities of assignment to treatment.  At each level of $\Gamma$, one finds the vector of unmeasured confounders for all individuals in the study, $\mathbf{u}$, which maximizes the $p$-value, hence yielding the worst possible inference for a given departure from purely random assignment. 

\begin{figure}[h]
\begin{center}
\begin{tikzpicture}[%
->,
shorten >=2pt,
>=stealth,
node distance=1cm,
pil/.style={
->,
thick,
shorten =2pt,},
rec/.style = {draw, rectangle}
]
\node [rec](2) {$U$};
\node[above left =of 2] (3) {$W_1$};
\node[above right =of 2](4) {$W_2$};
\node[above  =of 2](8) {$W_{12}$};
\node[below =of 2] (5) {$Z$};
\node[below left =of 5] (6) {$R_1$};
\node[below right =of 5] (7) {$R_2$};
\path 
(3) edge (2)
(4) edge (2)
(2) edge (5)
(3) edge (6)
(4) edge (7)
(5) edge (6)
(8) edge (2)
(8) edge [out = 150, in = 150] (6)
(8) edge [out = 30, in = 30 ] (7)
(5) edge (7);
\end{tikzpicture}
\caption{\small{A Directed Acyclic Graph (DAG) showing how our method accounts for unmeasured confounding on multiple outcome variables by controlling for their joint effect on the treatment. $W_1, W_2, W_{12}$ represent unmeasured factors which affect outcome $R_1$, outcome $R_2$, and both outcomes respectively. $U$ is an aggregate measure of the impact of $\{W_1, W_2, W_{12}\}$ on the treatment assignment vector, $Z$. For any known value of $U$, only the direct causal pathway of the treatment, $Z$, on the outcome variables, $R_1$ and $R_2$, remains open if we condition on $U$ (denoted by the square around $U$). Implicit in this diagram is that adjustment has been made for any observed pre-treatment confounders, $X$.}}
\label{fig:DAG}
\end{center}
\end{figure}
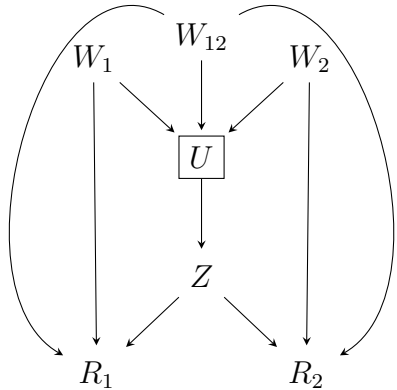

Matched observational studies may seek to investigate the impact of a single treatment on multiple outcome variables; see \citet{sab06}, \cite{voi12}, and \citet{obe14} for recent examples from policy analysis, economics, and health care. When there are multiple outcome variables of interest, there may exist unmeasured factors that influence a particular outcome while not impacting others. In order for these factors to affect the performed inference (and hence, to be confounders in the sense of \citet{van13}), these factors must also impact the treatment assignment probabilities. Figure \ref{fig:DAG} demonstrates that these factors yield an aggregate impact on the assignment probabilities ($U$ in the figure) despite affecting the outcomes differently. Controlling for the aggregate impact of unmeasured confounding on the assignment probabilities is sufficient for identifying the causal effect of the treatment on all of the outcome variables of interest, as these probabilities are themselves a minimally sufficient adjustment set \citep{ros83}.  The reader should keep in mind that $u_{ij}$ truly reflects a dimension reduction of all unmeasured confounders to their relevant scalar component for impacting the assignment probabilities, and hence that this model for a sensitivity analysis does not limit the potential impact of unmeasured confounding on any of the outcome variables. Moving forward, we will refer to $u_{ij}$ interchangeably as the ``unmeasured confounder" and ``unobserved covariate" for individual $j$ in matched set $i$,  as is conventional in sensitivity analyses following the model of \citet{obs}. 

When conducting a sensitivity analysis with multiple outcomes, the unmeasured confounder which affects assignment probabilities in the worst-case manner for outcome $k$, $\mathbf{u}^*_k$, may not be worst-case for outcome $k'$; in fact, it may actually result in more \textit{favorable} inference for outcome $k'$. As is noted in \citet{ros09}, na\"{\i}vely combining the results of outcome-specific sensitivity analyses while accounting for multiple comparisons is unduly conservative precisely because of this: it allows the worst-case unmeasured confounder to affect the probabilities of assignment to treatment differently from one outcome to the next for the same test subject.  Consequently, a uniform improvement in the power of a sensitivity analysis for testing the overall null hypothesis for any subset of outcomes could be attained by eliminating this logical inconsistency. As tests for the overall null hypothesis with respect to subsets of outcomes form the basis of multiple comparisons procedures such as closed testing \citep{mar76}, hierarchical testing \citep{mei08}, and the inheritance procedure \citep{goe12}, such an advance would also uniformly improve the power of a sensitivity analysis for testing null hypotheses for \textit{particular} outcomes while strongly controlling the familywise error rate.

The approaches for conducting a sensitivity analysis with a single outcome heavily utilize the fact that within each matched set, the search for the worst-case unmeasured confounder can be restricted to a readily enumerable set of binary vectors \citep{ros90}. When testing whether the treatment has an effect on \textit{at least one} of many outcome variables of interest, this is no longer the case. Thus, the potential gain in power cannot be actualized through simple extensions of existing methods. In this work, we present a new formulation of the required optimization problem as a quadratically constrained linear program which allows one to claim improved robustness to unmeasured confounding in an observational study with multiple outcomes when testing the overall null. This can, in turn, improve the reported robustness of individual level outcomes through its incorporation into certain sequential rejection procedures \citep{goe10}. To illustrate these ideas, we now present an observational study on the impact of smoking on naphthalene levels in the body.


\singlespacing
\subsection{Motivating Example: Naphthalene Exposure in Smokers}\label{sec:example}

Naphthalene is a simple polycyclic aromatic hydrocarbon (PAH) which has been linked to numerous adverse health outcomes. Exposure to excessive amounts of naphthalene can cause hemolysis (abnormal damage to or destruction of red blood cells in the body), which can in turn lead to hemolytic anemia \citep{tod91,san00}. Further, naphthalene has been shown to be carcinogenic in animal studies \citep{hec02}, prompting the International Agency for Research on Cancer (IARC) to label it as ``possibly carcinogenic to humans'' \citep{iarc02}. Given the potential for adverse health outcomes from exposure to naphthalene, it is of interest to assess the impact of other sources of exposure to naphthalene on levels of naphthalene metabolites found in the body.


 In the 2007-2008 National Health and Nutrition Examination Survey (NHANES), urinary concentrations of two monohydroxylated naphthalene metabolites, 1- and 2-naphthol (also known as $\alpha$- and $\beta$-naphthol) were collected for 1706 patients from a representative sample of adults aged 20 and older in the United States. Through this study, we seek to address the following question: after controlling for other sources of exposure and other relevant demographic variables, does smoking (one source of exposure to naphthalene) lead to elevated naphthalene metabolite levels in our study population? If this were the case, it would lend further credence to the belief that naphthalene is a useful biomarker for exposure to PAHs through inhalation \citep{nan01, hec02, pre04}, and it may serve to further highlight the health risks from smoking.

Through full matching \citep{han04}, 453 current smokers were placed into matched sets with 1253 non-current smokers who were similar on the basis of pre-treatment variables which, following the criterion for confounder selection of \citet{van11criterion}, were deemed important to the decision to be a smoker or the outcomes; see Appendix A for further details on the performed matching. Our two outcome variables were the urinary concentrations of 1- and 2-naphthol. Using an aligned rank test \citep{hod62} within the stratification yielded by our full match, we sought to determine whether there was evidence for smoking causing elevated levels for at least one of the two metabolites, and also whether smoking caused elevated metabolite levels for 1-naphthol and 2-naphthol considered individually. Assuming a multiplicative treatment effect model (additive on the log-scale), under no unmeasured confounding smoking was estimated to elevate urinary concentrations by a factor of 4.66 and 3.29 for 1- and 2-naphthol respectively using a Hodges-Lehman estimator \citep{hod63}, with 95\% confidence intervals of [4.00; 5.41] and [2.92; 3.69] attained by inverting a series of tests on the value of the multiplicative effect \citep{leh63}. Correcting for multiple comparisons using Holm-Bonferroni \citep{hol79}, the asymptotically separable algorithm of \citet{gas00} applied individually to each metabolite yielded strong insensitivity to unmeasured confounding: the minimum and maximum of the two outcome-specific findings were below 0.025 and 0.05 respectively until a $\Gamma$ of $7.78$. This means that an unmeasured confounder would have to result in a difference in the odds of smoking for two individuals in the same matched set by a factor of $7.78$ while nearly perfectly predicting naphthalene metabolite concentrations to render our results insignificant.

Based on these results, we can also attest to the robustness of a rejection of the \textit{overall} null of no effect for either naphthalene metabolite: we have evidence for significance of at least one naphthalene metabolite at $\Gamma=7.78$. As previously mentioned, this is conservative as using Holm-Bonferroni to combine individual sensitivity analyses allows for differing worst-case confounders for each outcome for the same individual. Naturally, the worst-case unmeasured confounder for 2-naphthol need not be the worst-case confounder for 1-naphthol. In fact, at $\Gamma=7.78$ the worst-case $\mathbf{u}$ for 2-naphthol actually yields a significant result for 1-naphthol, and similarly the worst-case $\mathbf{u}$ for 1-naphthol makes our result for 2-naphthol significant. Through the methodology presented in this paper, it can be determined  there is no vector of hidden covariates that simultaneously makes 1- and 2-naphthol insignificant at this level of unmeasured confounding. In fact, it takes a $\Gamma$ of $10.22$ to overturn the rejection of the overall null of no effect for either naphthalene metabolite. Thus $\Gamma=7.78$ actually understates the robustness of a test of overall significance. Furthermore, we show in Section \ref{sec:closed} that through a closed testing procedure we can actually claim robustness of the particular metabolites up until $\Gamma = 7.83$ for 1-naphthol and $\Gamma=8.20$ for 2-naphthol, which are the same levels of robustness to unmeasured confounding that would have been arrived upon \textit{without} controlling for multiple comparisons.

 Section \ref{sec:review} provides notation for and a review of randomization inference and sensitivity analysis within a matched observational study. Section \ref{sec:overall} introduces testing and sensitivity analysis for the overall null hypothesis when there are multiple outcomes. After highlighting the room for improvement relative to combining sensitivity analyses for each outcome, Section \ref{sec:improve} formulates a quadratically constrained linear program which allows us to perform a sensitivity analysis for the overall null hypothesis while enforcing that for each outcome, the unmeasured confounder must be the same for each individual. Section \ref{sec:closed} describes how our method can facilitate strong familywise error control for testing null hypotheses on particular outcomes through its incorporation into certain sequential rejection procedures. In Section \ref{sec:simulation}, we present a simulation study demonstrating the potential gains in power of a sensitivity analysis on the overall null and on outcome-specific nulls using this procedure. We return to our motivating example in Section \ref{sec:seq}, where we elucidate the improvements in reported robustness to unmeasured confounding attained through our procedure as they pertain to testing elevated naphthalene levels in smokers.

\singlespacing
\section{Notation for a Matched Observational Study}\label{sec:review}
\subsection{A Stratified Experiment with Multiple Outcomes}\label{sec:rand}
We now present notation for the idealized experiment targeted by matching algorithms wherein each treated unit is placed in a matched set with one or more control units. This framework can be trivially extended to dealing with strata resulting from full matching, such as the one presented in Section \ref{sec:example};  see \citet[][Section 4, Problem 12]{obs} for details. Suppose there are $I$ independent strata, the $i^{th}$ of which contains $n_i \geq 2$ individuals, that were formed on the basis of pre-treatment covariates. In each stratum, $ 1$ individual receives the treatment and $n_i-1$ individuals receive the control. There are $K$ outcome variables collected for each individual.  For each outcome $k$, individual $j$ in stratum $i$ has two potential outcomes: one under treatment, $r_{Tijk}$, and one under control, $r_{Cijk}$; let $\br_{Tij}$ and $\br_{Cij}$ be the $K$-dimensional vector of potential outcomes for this individual.  The observed response vector for each individual is $\bR_{ij} = \br_{Tij}Z_{ij} +\br_{Cij}(1-Z_{ij})$, where $Z_{ij}$ is an indicator variable that takes the value 1 if individual $j$ in stratum $i$ is assigned to the treatment; see, for example, \citet{ney23} and \citet{rub74}. Each individual has a vector of observed covariates $\x_{ij}$ and an unobserved covariate $u_{ij}$.

 There are $N = \sum_{i=1}^I n_i$ total individuals in the study. Let $\mathbf{Z} = [Z_{11}, Z_{12},...,$, $Z_{I n_{I}}]^T$ be the binary vector of treatment assignments, and let $\mathbf{R}, \br_{T}$, and $\br_{C}$ be the $N\times K$ matrices of observed responses and potential outcomes under treatment and control.  Let $\Omega$ be the set of $\prod_{i=1}^I n_i$ possible values of $\mathbf{Z}$ under the given stratification. In randomization inference for a randomized experiment, randomness is modeled solely through the assignment to treatment or to control \citep{fis35}. Quantities dependent on $\Z$, such as the observed outcomes $\mathbf{R}$, are random, while  $\br_{Tij}, \br_{Cij}, \x_{ij}$, and $u_{ij}$ are fixed across randomizations. Let $\mathcal{F}$ be the set of such fixed quantities. For a randomized experiment adhering to this design $\P(Z_{ij} = 1 | \cF, \Z \in \Omega) = 1/n_i$ and $\P(\Z = \z| \mathcal{F}, \Z \in \Omega) = 1/|\Omega|$, where $|A|$ denotes the number of elements in a finite set $A$. 

\singlespacing
\subsection{Randomization Inference and Sensitivity Analysis}\label{sec:sens}

For each outcome $k$, we consider hypotheses of the form $H_{k}: f_{Tk}(r_{Tijk}) = f_{Ck}(r_{Cijk})$ $\forall i,j$ for specified functions $f_{Tk}(\cdot)$ and $f_{Ck}(\cdot)$. For example, Fisher's sharp null of no effect can be tested through $f_{Tk}(r_{Tijk}) = r_{Tijk}$ and $f_{Ck}(r_{Cijk}) = r_{Cijk}$, and a test of an additive treatment effect $\tau_k$ can be tested by setting $f_{Tk}(r_{Tijk}) = r_{Tijk}$ and  $f_{Ck}(r_{Cijk}) = r_{Cijk} + \tau_k$. While tests for Neyman's weak null of no average treatment effect cannot be accommodated within the framework that follows, other choices of $f_{Tk}(\cdot)$ and $f_{Ck}(\cdot)$ can yield tests allowing for subject-specific causal effects such as tests of effect modification, dilated treatment effects, displacement effects, tobit effects, and attributable effects; see \citet[][Section 5]{obs} and \citet[][Sections 2.4-2.5]{designofobs} for an overview.

From our data alone we observe $F_{ijk} = f_{Tk}(r_{Tijk})Z_{ij} + f_{Ck}(r_{Cijk})(1-Z_{ij})$; let $\bF_k = [F_{11k},...,F_{In_{I}k}]$ .  Under $H_{k}$, the vectors $\bof_{Ck} = [f_{Ck}(r_{C11k}),...,f_{Ck}(r_{CIn_{I}k})]$ and $\bof_{Tk} = [f_{Tk}(r_{T11k}),...,$ $f_{Tk}(r_{TIn_{I}k})]$ are known to be equal, and hence are entirely specified. Further, they are constant across randomizations as they are known functions of the potential outcomes. Hence, under the null $\bF_k = \bof_{Tk} = \bof_{Ck} \in \mathcal{F}$, which in turn allows us to use randomization inference to test $H_{k}$. Specifically, under $H_{k}$ and under the stratified experiment discussed in Section \ref{sec:rand} the null distribution of a test statistic $t_{k}(\bZ, \bF_k)$ can be written as:
\begin{align}\label{eq:dist}\P\{t_{k}(\bZ, \bF_k) \geq a | \cF, \bZ \in\Omega; H_k\} = \frac{|\z \in \Omega: t_{k}(\z, \bof_{Ck}) \geq a|}{|\Omega|}, \end{align}
where we use $\bof_{Ck}$ in the right-hand side to emphasize that this distribution is known under the null.

The distribution of $t_{k}(\bZ, \bF_k)$ in (\ref{eq:dist}) is appropriate if the observed data truly resulted from the randomized experiment described in Section \ref{sec:rand}. In an observational study employing matching, we aim to replicate this idealized randomized experiment by creating strata wherein individuals are similar on the basis of their observed covariates, $\x_{ij}$ \citep{min00, han04, stu10}. While this seeks to control for observed confounders, individuals placed in a given stratum $i$ may be different on the basis of the unobserved covariate $u_{ij}$. If this $u_{ij}$ is influential for the assignment of treatments and the response, the distribution in (\ref{eq:dist}) may yield highly misleading inferences. 

 We follow the model for a sensitivity analysis discussed in \citet[Section 4]{obs}, which states that failure to account for unobserved covariates may result in biased treatment assignments within a stratum. This model can be parameterized by a number $\Gamma = \exp(\gamma) \geq 1$ which bounds the extent to which the odds ratio of assignment can vary between two individuals who are in the same matched stratum. Under this formulation, the probability of a given allocation of treatment and control within the stratification under consideration can be stated in the form
$\P(\mathbf{Z} = \z | \mathcal{F}, \bZ\in \Omega) = \exp(\gamma\z^T\bu)/\sum_{\bb \in \Omega} \exp(\gamma \bb^T\bu)$, where $\bu = [u_{11}, u_{12}, ..., u_{I, n_{i}}] \in [0,1]^N =: \mathcal{U}$ is a vector of unmeasured confounders for the individuals in the study. Note that $\Gamma = 1$ corresponds to the randomization distribution discussed in Section \ref{sec:rand}, while for $\Gamma > 1$ the resulting distribution differs from that of a randomized experiment, with $\Gamma$ controlling the extent of this departure.

 We consider test statistics of the form $t_k(\bZ, \bF_k) = \sum_{i=1}^I\sum_{j=1}^{n_i}Z_{ij}q_{ijk}$, where $q_{ijk}$ are functions of $\bF_k$. Under $H_k$ these values become functions of $\bof_{Ck}$, and hence are known quantities fixed across randomizations. Let $\bq_{k} = [q_{11k},...,q_{In_Ik}]$, and let  $\bq_{ik} = [q_{i1k},...,q_{in_ik}]$. Many commonly employed statistics can be written in this form. For example, suppose we are testing Fisher's sharp null, so that $F_{ijk} = R_{ijk}$, within the block-randomized experiment described in Section \ref{sec:rand}. Setting $q_{ijk} = \sum_{j'\neq j}( F_{ijk} - F_{ij'k})/(I(n_i-1))$, $t_k(\bZ, \bF_k)$ is the mean over the $I$ matched sets of the average treated-minus-control difference in each matched set for outcome $k$. In the case of a matched pairs design, $n_i = 2 \;\;\forall i$, this yields the paired permutation $t$-test. If $q_{ijk}$ are the ranks of the aligned response $F_{ijk} - \sum_{j'=1}^{n_i}F_{ij'k}/n_i$ from 1 to $N$, then a test on $t_k(\bZ, \bF_k)$ yields the aligned rank test of \citet{hod62}. To recover Wilcoxon's signed rank statistic for a matched pairs design, let $d_{ik}$ be the ranks of $|F_{i1k} - F_{i2k}|$ from 1 to I, and let $q_{ijk} = d_{ik}1\{F_{ijk} > F_{ij'k}\}$.  See \citet{obs} for additional examples and further discussion. 
 
 For any given value of $\Gamma \geq 1$, a sensitivity analysis proceeds by finding the allocation of the unmeasured confounder $\bu^*$ which maximizes the $p$-value for the hypothesis test being conducted. While not explicitly noted, this worst-case unmeasured confounder can vary with the value of $\Gamma$ under investigation. One then finds the smallest value of $\Gamma$ such that the conclusions of the study would be altered (i.e., such that the conclusion of the hypothesis test would change from rejecting to failing to reject the null hypothesis). The more robust a given study is to unmeasured confounding, the larger the value of $\Gamma$ must be to alter its findings. Under mild regularity conditions on $\bq_k$, the distribution under the null of $t_{k}(\bZ, \bF_k)$ converges to that of a normal random variable as $I\rightarrow \infty$ for the worst-case confounder $\bu^*$ at any $\Gamma$. An example of regularity conditions on the constants $q_{ijk}$ is that the Lindeberg condition holds for the random variables $B_{ik} := \sum_{j=1}^{n_i}Z_{ij}q_{ijk}$ \citep[][Theorem A.1.1]{leh04}. While the value of $\Gamma$ itself does not affect the limiting distribution, it does influence the rate at which this limit is reached as larger values of $\Gamma$ allow for larger discrepancies in the assignment probabilities within a matched set. Under asymptotic normality, large sample bounds on the tail probability can instead be expressed in terms of corresponding bounds on standardized deviates.

 For further discussion of sensitivity analyses, including illustrations and alternate models, see \citet{cor59}, \citet{mar97}, \citet{imb03}, \citet{yu05}, \citet{wan06}, \citet{egl09}, \citet{hos10}, \citet{van11}, \citet{zub13}, \citet{liu13} and \citet{din14corn}.

\section{Sensitivity Analysis for Overall Significance}\label{sec:overall}
\subsection{Testing the Overall Null Hypothesis}
We begin with notation for the truth of the null hypotheses on all $K$ outcomes; extensions of notation to dealing with subsets of outcomes, which will in turn facilitate strong familywise error control for testing individual outcomes, will be made in Section \ref{sec:closed}.  There are $K$ hypotheses, $H_1,...,H_K$, and we are interested in testing the overall truth of the hypotheses $\{H_1,..,H_K\}$ while strongly controlling the familywise error rate at level $\alpha$ for a range of $\Gamma$.\begin{align*} \mathbf{H_o}: \bigwedge_{k=1}^{K} H_{k}\\
\mathbf{H_a}: \bigvee_{k=1}^{K} H_{k}^c
\end{align*} We will refer to a test of $\mathbf{H_o}$ as a test of the \textit{overall null}. Moving forward, we assume each individual hypothesis $H_k$ has an associated test statistic $t_{k}(\bZ, \bF_k)$ of the form discussed in Section \ref{sec:sens}. 

\singlespacing
\subsection{Combining Individual Sensitivity Analyses is Conservative}

A simple approach for conducting a sensitivity analysis at a given $\Gamma$ would be to separately find the worst-case $p$-value for each hypothesis test, call it ${P}^*_k$ with corresponding allocation of worst-case confounder ${\bu}^*_k$, and suggest through the use of a Bonferroni correction that at least one hypothesis is false if $\min_{k} {P}^*_k \leq \alpha/K$. This trivially controls familywise error rate at $\alpha$ as desired; however, as is noted in \citet[][Section 4.5]{ros09}, this approach is conservative as the worst-case $p$-value for hypothesis test $k$ may be found at a different allocation of the unmeasured confounder as that of hypothesis test $k' \neq k$  for $k, k' \in \{1,...,K\}$ (i.e., ${\bu}^*_k \neq {\bu}^*_{k'}$). In other words, the biased treatment assignment probabilities caused by unmeasured confounding that yield the worst-case inference for outcome $k$ and outcome $k'$ need not be the same. This can be better understood in light of the following well known minimax inequality \citep[for instance,][Lemma 1.3.1]{kar92}
\begin{align}\label{eq:minmax}
 \min_{k \in \{1,..,K\}}\;\max_{\bu \in \mathcal{U}}  P_{k,\bu} \geq \max_{\bu \in \mathcal{U}}\;  \min_{k \in \{1,..,K\}}\; P_{k,\bu}.
\end{align}Combining the results of $K$ separate hypothesis tests and Bonferroni correcting corresponds to the left-hand side of (\ref{eq:minmax}). Strict inequality is possible in (\ref{eq:minmax}): it could be the case that $\min_k\max_{\bu \in \mathcal{U}} P_k > \alpha/K$, meaning that we would fail to reject the overall null hypothesis if we conducted sensitivity analyses separately for each $k$ and then Bonferroni corrected, while in reality $\max_{\bu \in \mathcal{U}} \min_k P_k \leq \alpha/K$, such that we should have rejected the overall null. This would occur if for each $k$ there exists a $\bu^*_k \in \mathcal{U}$ such that $H_k$ is not rejected, yet there does not exist a \textit{single} $\mathbf{u}^* \in \mathcal{U}$ for which all $H_k$ are simultaneously not rejected. 

 A uniform improvement over combining individual sensitivity analyses could be achieved by a procedure which directly solved for the right-hand side of (\ref{eq:minmax}). Such a procedure cannot be derived by extending existing methods for conducting individual level sensitivity analyses, as these methods rely upon the fact that the search for a worst-case confounder can be restricted to vectors in $\mathcal{U}^+$ or $\mathcal{U}^-$ for any particular hypothesis $k$. Unfortunately, it is not the case that vector $\bu^*$ which achieves \\$\max_{\bu \in \mathcal{U}} \min_{k \in \{1,..,K\}}\; P_{k,\bu}$ lies within an easily enumerated set of vertices of $\mathcal{U}$; in fact, the solution need not even lie at a vertex. To exploit this potential improvement, a new formulation of the required optimization problem that allows for solutions in all of $\mathcal{U}$ is thus required.

 \singlespacing
\section{Improving Power through Quadratically Constrained Linear Programming}\label{sec:improve}

In this section, we assume the individual level hypotheses $H_k$ have two-sided alternatives; simple extensions to the one-sided case are discussed in Appendix B. Using a normal approximation, we can equivalently express our problem as minimizing over $\mathcal{U}$ the maximal squared deviate over the $K$ hypotheses in question:
\begin{align}\label{eq:maxsquare}
\min_{\bu \in \mathcal{U}}\;  \max_{k \in \{1,..,K\}}\; \frac{(t_k - \mu_{k,\bu})^2}{\sigma^2_{k,\bu}},
\end{align} where $t_k$ is the observed value of the statistic $t_{k}(\bZ, \bF_k)$, and $\mu_{k,\bu} = \E_{\Gamma,\bu}[\Z^T\bq_k|\cF, \bZ\in\Omega]$ and $\sigma^2_{k,\bu} =  \Var_{\Gamma,\bu}(\Z^T\bq_k|\cF, \bZ\in\Omega)$ are the means and variances of the test statistic $t_{k}(\bZ, \bF_k)$ with a given value of $\Gamma$ and vector $\bu$ under the permutation distribution given by (\ref{eq:dist}). Under a normal approximation for $t_{k}(\bZ, \bF_k)$, the squared deviate follows a $\chi^2_1$ distribution. Hence, a determination of whether or not we can reject at least one null hypothesis can be made by checking whether or not the solution to (\ref{eq:maxsquare}) is  greater than or equal to $\chi^2_{1, 1-\alpha/K}$, where $\chi^2_{1,1-\alpha/K}$ is the $1-\alpha/K$ quantile of a $\chi^2_1$ distribution.

Moving forward, all expectations and variances are taken with respect to the distribution in (\ref{eq:dist}), i.e. under the truth of the null hypothesis $H_k$ for each $k$, and are conditional on $\mathcal{F}$ and $\bZ\in\Omega$; this is omitted for notational ease. Let $\varrho_{ij} =  \exp(\gamma u_{ij}) /\sum_{j'=1}^{n_i}\exp(\gamma u_{ij'}) = \P(Z_{ij}=1 | \mathcal{F}, \bZ \in \Omega)$. Let $\brho_i = [\varrho_{i1},..,\varrho_{in_i}]$, and let $\brho = [\varrho_{11},..,\varrho_{In_I}]$. Note that we can express our test statistics as the sums of stratum-wise contributions, $t_k(\bZ, \bF_k) = \sum_{i=1}^IB_{ik}$ where $B_{ik} := \sum_{j=1}^{n_i}Z_{ij}q_{ijk}$. The expectation and variance of the contribution from stratum $i$, $ B_{ik}$, can be written as\begin{align*}
\E[B_{ik};\brho] &= \brho_i^T\bq_{ik}\\
\Var (B_{ik};\brho) &= \brho_i^T {\bq}_{ik}^2 - (\brho_i^T{\bq}_{ik})^2,
\end{align*} where the simplified form of $\Var(B_{ik};\brho)$ comes from the constraint that $\sum_{j=1}^{n_i}Z_{ij} = 1 \; \forall i$. 

For a given $\brho$, we can reject the null hypothesis for a two sided alternative at level $\alpha/K$ if 
$(t_k - \E[t_{k}(\bZ, \bF_k);\brho])^2/\Var(t_{k}(\bZ, \bF_k);\brho) \geq \chi^2_{1,1-\alpha/K}$, where $\E[t_{k}(\bZ, \bF_k);\brho] = \sum_{i=1}^I\E[B_{ik};\brho]$, and $\Var(t_{k}(\bZ, \bF_k); \brho) = \sum_{i=1}^I \Var(B_{ik};\brho)$ due to independence between strata. This is equivalent to rejecting if  $\zeta_k(\brho) := (t_k - \E[t_{k}(\bZ, \bF_k);\brho])^2 -  \chi^2_{1,1-\alpha/K}\Var(t_{k}(\bZ, \bF_k);\brho) \geq 0$. If we can determine that $\zeta_k(\brho) \geq 0$ for all feasible values of $\brho$ at a given value of $\Gamma$, we can then say that we have rejected the null at level of unmeasured confounding $\Gamma$; otherwise, we fail to reject. 

 Consider the following optimization problem:
\begin{align*}
\tag{$H_k$}
\label{eq:simple}
\underset{\varrho_{ij}, s_i}{\text{minimize}} &\;\;\zeta_k(\brho) \\
\text{subject to} & \;\; \sum_{j=1}^{n_i}\varrho_{ij} = 1 \;\;\forall i \nonumber\\
&s_i\leq \varrho_{ij} \leq \Gamma s_i\;\; \forall i,j\\
&\varrho_{ij} \geq 0 \;\; \forall i,j
\end{align*} 


The variables $s_i$ stem from an application of a Charnes-Cooper transformation, $s_i = 1/ \sum_{j'=1}^{n_i}\exp(\gamma u_{ij'})$ \citep{cha62}, and allow us to incorporate the restrictions on the allowable departure from pure randomization, $1 \leq \exp(\gamma u_{ij}) \leq \Gamma \;\; \forall i,j$, in terms of the probabilities themselves. 

Problem (\ref{eq:simple}) is a quadratic program, which can be readily solved using a host of free and commercially available solvers; however, solving this problem merely results in a sensitivity analysis for a \textit{particular} hypothesis $H_k$, rather than one of the overall null $\wedge H_k$. Towards this end, define $\zeta(\brho) = \max \{\zeta_1(\brho),...,\zeta_K(\brho)\}$. We can now pose our problem as finding $\min_\brho \zeta (\brho)$ subject to constraints on $\brho$ imposed by $\Gamma$. This optimization can be performed through incorporating an auxiliary variable $y$ and solving the following quadratically constrained linear program:
\begin{align*}
\tag{$\wedge H_k$}
\label{eq:wedge}
\underset{{y,\varrho_{ij},s_i} }{\text{minimize}} &\;\; y\\
\text{subject to}\;\; & y \geq \zeta_k(\brho)\;\; \forall k\\
& \;\; \sum_{j=1}^{n_i}\varrho_{ij} = 1 \;\;\forall i \nonumber\\
&s_i\leq \varrho_{ij} \leq \Gamma s_i\;\; \forall i,j\\
&\varrho_{ij} \geq 0 \;\; \forall i,j
\end{align*}

The auxiliary variable $y$ is forced to be larger than $\zeta_k(\brho)$ for all $k$, and by minimizing over $y$ the optimization problem searches for the feasible value of $\brho$ that allows for $y$ to become as small as possible, hence minimizing the maximum value as desired. This is a commonly employed device for solving minimax problems; see, for example, \citet{cha78}. To determine whether or not we can reject at least one null hypothesis, we simply check whether the optimal value $y^*\geq 0$. If it is, we can reject at least one null hypothesis; otherwise, we cannot. Quadratically constrained linear programs can be solved using many available solvers; we provide an implementation using the \texttt{R} interface to \texttt{Gurobi}, a commercial solver which is freely available for academic use. Henceforth, we will refer to this procedure for conducting a sensitivity analysis the overall null with $K$ outcomes as the ``minimax'' procedure (for minimizing the maximum squared deviate).

\singlespacing
\section{Familywise Error Control for Individual Null Hypotheses}\label{sec:closed}

By addressing the right-hand side of (\ref{eq:minmax}), the minimax procedure provides a sensitivity analysis for the overall null hypothesis that uniformly dominates combining individual sensitivity analyses. In this section, we discuss how the minimax procedure can be used with sequential rejection procedures \citep{goe10} which progress through testing the overall null for a sequence of subsets of outcomes (henceforth referred to as intersection nulls) to provide uniform improvements in power for testing hypotheses on \textit{particular} outcome variables. Sequential rejection procedures of this sort include closed testing \citep{mar76}, hierarchical testing \citep{mei08}, and the inheritance procedure \citep{goe12}. These procedures have appealing properties for conducting a sensitivity analysis, often allowing researchers to claim improved robustness of a study's findings against unmeasured confounding; see \citet{ros09} for a discussion of this fact as it relates to closed testing procedures. 

We now introduce notation for the class of sequential rejection procedures which can be used in conjunction with our method, i.e. those for which each step involves testing the truth of an intersection null hypothesis for a subset of the $K$ outcome variables. There are $L$ intersection null hypotheses ordered from 1,...,$L$, the $\ell^{th}$ of which, $\mathbf{H_o}_\ell$, pertains to the null hypothesis being true for all outcomes in the subset $\mathcal{K}_\ell \subseteq \{1,...,K\}$. That is, $\mathbf{H_{o\ell}} = \bigwedge_{k  \in \mathcal{K}_\ell}H_{k\ell}$. $|\mathcal{K}_\ell| \leq K$ is the number of outcomes being tested in the $\ell^{th}$ subset; $|\mathcal{K}_\ell| = 1$ then corresponds to a test of a particular outcome.  Let $\mathcal{H}$ be the set of these $L$ intersection null hypotheses, $\mathcal{H} = \{\mathbf{H_o}_1,..., \mathbf{H_o}_L\}$.

Following \citet{goe10}, let $\mathcal{R}_a \subseteq \mathcal{H}$ be the collection of intersection nulls rejected after step $a$ of the sequential rejection procedure, and let $\mathcal{N}(\mathcal{R}_a)$ be the set of intersection nulls that can now be rejected in step $a+1$ if all elements of $\mathcal{R}_a$ have been rejected by step $a$. The sequential rejection procedure can then be defined by \begin{align*}\mathcal{R}_0 &= \emptyset\\
\mathcal{R}_{a+1} &= \mathcal{R}_a \cup \mathcal{N}(\mathcal{R}_a),\end{align*}
and is repeated until convergence (i.e., until $\mathcal{R}_{a+1} = \mathcal{R}_a$).  \citet{goe10} show that sequential rejection procedures strongly control the familywise error rate at $\alpha$ under the conditions (1) the procedure controls the familywise error at $\alpha$ for the so-called \textit{critical case} in which procedure has rejected all of the false overall null hypotheses and none of the true overall nulls and (2) no false rejections in the critical case implies no false rejections in situations with fewer rejections than the critical case. 

Closed testing, hierarchical testing, and the inheritance procedure can all be recovered through specific choices of $\mathcal{N}(\cdot)$ that provably adhere to these conditions. Testing the intersection nulls $\mathbf{H_o}_\ell$ for any $\ell$ at level of unmeasured confounding $\Gamma$ as required by these procedures can be performed using the minimax procedure of Section \ref{sec:improve}, which through inequality (\ref{eq:minmax}) provides improved power for each subset tested. 

To illustrate, suppose one is interested in using a closed testing procedure to conduct a sensitivity analysis with $K=2$ outcomes; this is the procedure used for multiple testing in our motivating example. In this case, $L=3$, $\mathcal{K}_1 = \{1,2\}$, $\mathcal{K}_2 = \{1\}$, $\mathcal{K}_3 = \{2\}$. The function $\mathcal{N}(\cdot)$ then takes on the following form:
\begin{align*}
\mathcal{N}(\emptyset) &=\begin{cases} \mathbf{H_o}_1& \text{if  reject } H_1\wedge H_2 \text{ at level }\alpha\\ \emptyset & \text{ otherwise} \end{cases}\\
\mathcal{N}(\mathbf{H_o}_1) &= \begin{cases}  \{\mathbf{H_o}_1, \mathbf{H_o}_2, \mathbf{H_o}_3\} & \text{ if } H_1 \text{ and } H_2 \text{ each reject individually at level } \alpha\\
 \{\mathbf{H_o}_1, \mathbf{H_o}_2\} & \text{ if only } H_1 \text{ rejects at level } \alpha\\
  \{\mathbf{H_o}_1, \mathbf{H_o}_3\} & \text{ if only } H_2 \text{ rejects at level } \alpha\\
   \{\mathbf{H_o}_1\} & \text{otherwise}, \end{cases}\end{align*} and $\mathcal{N}(A) = A$ if $A \neq \emptyset$ and $A \neq \mathbf{H_o}_1$.  In this example, the test of $\mathbf{H_o}_1$ can be performed using the minimax procedure with a test that is locally level $\alpha$; the tests of $\mathbf{H_o}_2$ and $\mathbf{H_o}_3$ only involve one outcome and thus can be conducted through the usual methods for a sensitivity analysis which, by the closure principle, can be performed locally at level $\alpha$ while strongly controlling the familywise error rate. 

\singlespacing
\section{Simulation Study: Gains in Power of a Sensitivity Analysis}\label{sec:simulation}
\subsection{Overall Null Hypothesis}
Through the minimax procedure, we arrive at a uniform improvement for testing the overall null relative to combining the results of individual sensitivity analyses. In this section, we present a simulation study to demonstrate the potential gains in power for testing the overall null. In each of 24 simulation settings, we simulate 10,000 data sets with $I=250$ pairs and $K=5$ outcome variables of interest. The vector of treated-minus-control paired differences $\mathbf{D}_i$ are simulated $iid$ from a multivariate normal with mean vector $\btau$ and covariance matrix $\Sigma$. For each outcome, we use an M-statistic of the type favored by \citet{hub81}, $t_k(\bZ, \bF_k) = \sum_{i=1}^I\psi(D_{ik}/s_k)$, to conduct inference, where $s_k$ is the median of $|D_{ik}|$ across individuals $i$ and $\psi(y) = \text{sign}(y)\min(|y|, 2.5)$. See \citet{mar79} for a discussion of randomization inference for $M$-statistics, and see \citet{ros07,ros13, ros14} for various aspects of sensitivity analyses for $M$-statistics.

 In evaluating these two procedures, we assume as is advocated in \citet{ros04, ros07} that unbeknownst to the practitioner the paired data at hand truly arose from a stratified randomized experiment (i.e., $\Gamma=1$). Hence, using a standard randomization test without assuming unmeasured confounding would provide honest type I error control. The practitioner, blind to this, would like to not only perform inference under the assumption of no unmeasured confounding, but also assess the robustness of the study's findings to unobserved biases of varying severity. 
 
 \begin{table}
\caption{ \label{tab:overturn}Power of a sensitivity analysis for the overall null. }
 \begin{center}
\begin{tabular}{c  c c c}
\hline
Gamma&Moments&Separate & Minimax\\
\hline
\multirow{8}{*}{$\Gamma = 1.25$}&	$\btau^{(1)},	\Sigma^{(1)}$&	0.94	&	0.99	\\
&	$\btau^{(1)},	\Sigma^{(2)}$&	0.77	&	0.80	\\
&	$\btau^{(2)},	\Sigma^{(1)}$&	0.89	&	0.96	\\
&	$\btau^{(2)},	\Sigma^{(2)}$&	0.73	&	0.77	\\
&	$\btau^{(3)},	\Sigma^{(1)}$&	0.92	&	0.96	\\
&	$\btau^{(3)},	\Sigma^{(2)}$&	0.85	&	0.87	\\
&	$\btau^{(4)},	\Sigma^{(1)}$&	0.72	&	0.72	\\
&	$\btau^{(4)},	\Sigma^{(2)}$&	0.71	&	0.72	\\
\hline						
\multirow{8}{*}{$\Gamma = 1.5$}&	$\btau^{(1)},	\Sigma^{(1)}$&	0.34	&	0.78	\\
&	$\btau^{(1)},	\Sigma^{(2)}$&	0.25	&	0.33	\\
&	$\btau^{(2)},	\Sigma^{(1)}$&	0.28	&	0.66	\\
&	$\btau^{(2)},	\Sigma^{(2)}$&	0.21	&	0.28	\\
&	$\btau^{(3)},	\Sigma^{(1)}$&	0.45	&	0.65	\\
&	$\btau^{(3)},	\Sigma^{(2)}$&	0.39	&	0.45	\\
&	$\btau^{(4)},	\Sigma^{(1)}$&	0.26	&	0.26	\\
&	$\btau^{(4)},	\Sigma^{(2)}$&	0.25	&	0.25	\\
\hline						
						
\multirow{8}{*}{$\Gamma = 1.75$}&	$\btau^{(1)},	\Sigma^{(1)}$&	0.04	&	0.36	\\
&	$\btau^{(1)},	\Sigma^{(2)}$&	0.03	&	0.06	\\
&	$\btau^{(2)},	\Sigma^{(1)}$&	0.03	&	0.23	\\
&	$\btau^{(2)},	\Sigma^{(2)}$&	0.03	&	0.05	\\
&	$\btau^{(3)},	\Sigma^{(1)}$&	0.09	&	0.24	\\
&	$\btau^{(3)},	\Sigma^{(2)}$&	0.09	&	0.12	\\
&	$\btau^{(4)},	\Sigma^{(1)}$&	0.05	&	0.05	\\
&	$\btau^{(4)},	\Sigma^{(2)}$&	0.04	&	0.04	\\
\hline

 \end{tabular}
  \end{center}
 \end{table}

 Our 24 simulation settings are the $8$ possible combinations of the following mean and covariance vectors, each tested at $\Gamma=1.25, 1.5$ and $1.75$:
\begin{enumerate}
\item $\btau^{(1)} = [0.25,0.25,0.25,0.25,0.25]$; $\btau^{(2)} = [0.25,0.25,0.25,0.25,0]$;\\  $\btau^{(3)} = [0.3,0.3, 0, 0, 0]$; $\btau^{(4)} = [0.3,0, 0, 0, 0]$ 
\item $\Sigma^{(1)} = \text{Diag}(1)$; $\Sigma^{(2)}_{ij} = 1$ if $i=j$, $\Sigma^{(2)}_{ij} = 0.5$ otherwise.
 \end{enumerate}

%
%
%

 All hypothesis tests are of Fisher's sharp null, and are conducted with two-sided alternatives at $\alpha = 0.05$. Table \ref{tab:overturn} displays the probabilities of (correctly) rejecting the overall null of no effect for any of the outcomes. The first column contains the probabilities of rejection when combining the results of individual sensitivity analyses, while the second contains these probabilities for the minimax procedure. The relative improvement through the minimax procedure can be quite substantial when the paired differences are independent across outcomes $(\Sigma^{(1)})$, while more modest improvements are attained when the paired differences are positively correlated $(\Sigma^{(2)})$. With positively correlated differences across outcomes, the worst-case unmeasured confounder for a particular outcome begins to align more closely with the worst-case unmeasured confounder for the other outcomes, while for independent paired differences this often is not the case. For both independent and correlated paired differences, gains are also more substantial when there are 5 or 4 nonzero treatment effects $(\btau^{(1)}$ and $\btau^{(2)}$) versus 2 larger nonzero effects ($\btau^{(3)}$), and with only one large nonzero effects ($\tau^{(4)}$) the two methods tend to coincide. With fewer nonzero effects, the significance of the overall null at a given level of unmeasured confounding depends on the pattern of paired differences in a small number of outcomes, such that even if the worst-case unmeasured confounder for an outcome with a nonzero effect actually improves the squared deviate for an outcome with zero effect it is unlikely to elevate said deviate to a level of significance.
 
 Naturally, the probabilities of rejection decrease as $\Gamma$ increases for each combination of mean vector and covariance matrix. We also note that as $\Gamma$ increases, the gains from using the minimax procedure also increase . For example, with combination $\btau^{(2)}, \Sigma^{(1)}$ the powers of the combined approach versus the minimax approach are 0.89 and 0.96 at $\Gamma=1.25$, and are 0.28 versus 0.66 at $\Gamma=1.5$. These simulations indicate that conducting a sensitivity analysis for the overall null by minimizing the maximum squared deviate allows for substantial and clinically relevant gains in the power of a sensitivity analysis. Additionally, the computational burden of the required optimization problem was minimal in these simulations: across all 24 simulation settings, the average computation time on a desktop computer with a 3.40 GHz processor and 16.0 GB RAM was 0.12 seconds.

\subsection{Individual Hypotheses}\label{sec:holm}
As discussed in Section \ref{sec:closed}, the benefits of our procedure extend beyond testing the overall null, and can in fact yield improved power for a sensitivity analysis on hypotheses for individual outcomes. To illustrate this fact, we present a simulation study assessing the individual-level power of a sensitivity analysis for each of $K=3$ outcomes. We use a closed testing procedure in order to test hypotheses on individual outcomes. Briefly, the closed testing principle states that if there are $K$ hypotheses $H_1,...,H_K$ that are of interest, we can reject any particular hypothesis $H_k$ with familywise error control at $\alpha$ if all intersections of hypotheses including $H_k$ can be rejected with tests that are individually level $\alpha$. For example, with three outcomes we can reject $H_1$ if we can reject $H_1\wedge H_2 \wedge H_3$, $H_1\wedge H_2$, $H_1 \wedge H_3$, and $H_1$ with tests that are locally level $\alpha$. When combining the results of individual sensitivity analyses, this equates to the Holm-Bonferroni procedure. When using the minimax procedure for closed testing, one instead solves problem (\ref{eq:wedge}) for each intersection hypothesis.

  \begin{table}
  \begin{center}
\caption{ \label{tab:individ}Power of closed testing for individual nulls.}
\begin{tabular}{c c | c c c c | c c c c}
\hline
&\multicolumn{1}{c}{}&\multicolumn{4}{c}{Separate} & \multicolumn{4}{c}{Minimax}\\
\multicolumn{1}{c}{Gamma} & \multicolumn{1}{c}{Moments} & $H_1$ & $H_2$ &$H_3$& \multicolumn{1}{c}{$\wedge H_k$} & $H_1$ & $H_2$ & $H_3$ &$\wedge H_k$\\
\hline
\multirow{8}{*}{$\Gamma = 1.25$}&	$\btau^{(1)},	\Sigma^{(1)}$&	0.27	&	0.40	&	0.54	&	0.74	&	0.33	&	0.46	&	0.60	&	0.84	\\
&	$\btau^{(1)},	\Sigma^{(2)}$&	0.29	&	0.40	&	0.53	&	0.62	&	0.31	&	0.43	&	0.56	&	0.65	\\
&	$\btau^{(2)},	\Sigma^{(1)}$&	0.65	&	0.86	&	0.96	&	0.99	&	0.68	&	0.88	&	0.97	&	1.00	\\
&	$\btau^{(2)},	\Sigma^{(2)}$&	0.65	&	0.85	&	0.95	&	0.97	&	0.66	&	0.86	&	0.96	&	0.97	\\
&	$\btau^{(3)},	\Sigma^{(1)}$&	0.32	&	0.59	&	0.95	&	0.97	&	0.35	&	0.63	&	0.97	&	0.99	\\
&	$\btau^{(3)},	\Sigma^{(2)}$&	0.34	&	0.58	&	0.94	&	0.95	&	0.35	&	0.60	&	0.95	&	0.95	\\
&	$\btau^{(4)},	\Sigma^{(1)}$&	0.09	&	0.27	&	0.94	&	0.95	&	0.11	&	0.29	&	0.95	&	0.97	\\
&	$\btau^{(4)},	\Sigma^{(2)}$&	0.11	&	0.27	&	0.93	&	0.94	&	0.11	&	0.28	&	0.94	&	0.94	\\
\hline																		
\multirow{8}{*}{$\Gamma = 1.375$}&	$\btau^{(1)},	\Sigma^{(1)}$&	0.09	&	0.16	&	0.27	&	0.41	&	0.14	&	0.22	&	0.34	&	0.61	\\
&	$\btau^{(1)},	\Sigma^{(2)}$&	0.11	&	0.18	&	0.27	&	0.35	&	0.13	&	0.20	&	0.30	&	0.39	\\
&	$\btau^{(2)},	\Sigma^{(1)}$&	0.37	&	0.63	&	0.85	&	0.94	&	0.42	&	0.70	&	0.90	&	0.99	\\
&	$\btau^{(2)},	\Sigma^{(2)}$&	0.39	&	0.62	&	0.84	&	0.87	&	0.41	&	0.65	&	0.85	&	0.89	\\
&	$\btau^{(3)},	\Sigma^{(1)}$&	0.12	&	0.31	&	0.83	&	0.87	&	0.16	&	0.37	&	0.88	&	0.95	\\
&	$\btau^{(3)},	\Sigma^{(2)}$&	0.14	&	0.32	&	0.82	&	0.83	&	0.16	&	0.35	&	0.83	&	0.84	\\
&	$\btau^{(4)},	\Sigma^{(1)}$&	0.02	&	0.10	&	0.81	&	0.83	&	0.03	&	0.12	&	0.85	&	0.89	\\
&	$\btau^{(4)},	\Sigma^{(2)}$&	0.03	&	0.11	&	0.82	&	0.82	&	0.03	&	0.12	&	0.82	&	0.82	\\
\hline																		
\multirow{8}{*}{$\Gamma = 1.5$}&	$\btau^{(1)},	\Sigma^{(1)}$&	0.03	&	0.06	&	0.11	&	0.18	&	0.05	&	0.09	&	0.16	&	0.36	\\
&	$\btau^{(1)},	\Sigma^{(2)}$&	0.03	&	0.06	&	0.12	&	0.16	&	0.05	&	0.08	&	0.14	&	0.19	\\
&	$\btau^{(2)},	\Sigma^{(1)}$&	0.16	&	0.38	&	0.64	&	0.77	&	0.22	&	0.48	&	0.76	&	0.95	\\
&	$\btau^{(2)},	\Sigma^{(2)}$&	0.18	&	0.38	&	0.64	&	0.69	&	0.20	&	0.42	&	0.68	&	0.74	\\
&	$\btau^{(3)},	\Sigma^{(1)}$&	0.04	&	0.13	&	0.62	&	0.66	&	0.06	&	0.18	&	0.71	&	0.84	\\
&	$\btau^{(3)},	\Sigma^{(2)}$&	0.04	&	0.14	&	0.62	&	0.63	&	0.05	&	0.16	&	0.64	&	0.66	\\
&	$\btau^{(4)},	\Sigma^{(1)}$&	0.00	&	0.03	&	0.62	&	0.63	&	0.01	&	0.04	&	0.67	&	0.73	\\
&	$\btau^{(4)},	\Sigma^{(2)}$&	0.01	&	0.04	&	0.62	&	0.62	&	0.01	&	0.04	&	0.63	&	0.63	\\
\hline
 \end{tabular}
  \end{center}
 \end{table}

 In each of 24 simulation settings, we simulate 10,000 data sets under no unmeasured confounding with $I=250$ pairs for the three outcome variables of interest and again use Huber's M-statistic. For each of the 8 combinations of treatment effects and covariances, closed testing is used to test individual hypotheses, and tests are run at $\Gamma = 1.25, 1.375$, and $1.5$. We also include the power for rejecting the overall null for each combination and at each level of $\Gamma$. The values for the treatment effect vector and the covariances were as follows:
\begin{enumerate}
\item $\btau^{(1)} = [0.2, 0.225, 0.25]$; $\btau^{(2)} = [0.25,0.3,0.35]$;  $\btau^{(3)} = [0.2,0.25,0.35]$;\\ $\btau^{(4)} =  [0.15,0.25,0.35]$
\item $\Sigma^{(1)} = \text{Diag}(1)$; $\Sigma^{(2)}_{ij} = 1$ if $i=j$, $\Sigma^{(2)}_{ij} = 0.5$ otherwise.
 \end{enumerate}

 Table \ref{tab:individ} shows the power for rejecting Fisher's sharp null for each outcome under four different vectors of true treatment effect values and two different forms of the covariance matrix. The magnitude of the improvement attained through the minimax procedure can be seen to depend on many factors. All else equal, as $\Gamma$ increases the gains in power also increase. The gains in power tend to be more substantial in the $iid$ cases ($\Sigma^{(1)}$) versus the positively correlated case $(\Sigma^{(2)}$), as for each intersection hypothesis the minimax procedure tends to resemble more closely the individual testing approach when there is positive correlation since the worst-case confounders across outcomes tend to align more closely. For example, with $\btau^{(2)} = [0.25, 0.3, 0.35]$ at $\Gamma=1.5$, the power after combining individual sensitivity analyses and after using the minimax procedure are $[0.16, 0.38, 0.64]$ versus $[0.22, 0.48, 0.76]$  when the paired differences are independent across outcomes, yet were $[0.18, 0.38, 0.64]$ versus $[0.20, 0.42, 0.68]$ when positively correlated. Gains are also most apparent when the treatment effects are of roughly the same magnitude ($\btau^{(1)}$ and $\btau^{(2)}$), while the gains tail off as one outcome increasingly determines the rejection of the overall null (compare $\btau^{(2)}, \btau^{(3)}, \btau^{(4)})$. Thus, while the gains for testing the overall null hypothesis may be most apparent, the minimax procedure can provide meaningful improvements for testing nulls on individual outcomes.
 
 In Appendix C, we show that our procedure does provide strong familywise error control in the presence of true intersection nulls as desired.

\singlespacing
\section{Improved Robustness to Unmeasured Confounding for Elevated Napthalene in Smokers}\label{sec:seq}
\subsection{Conflicting Desires for the Worst-Case Confounder}
%
%

\begin{table}[h]
\begin{center}
\caption{\label{tab:pair} Worst-Case Confounders in a Particular Pair at $\Gamma = 10$}
\vspace{.1 in}
\begin{tabular}{|c| c c c c | c c c c|}
\multicolumn{1}{c}{}&\multicolumn{4}{c}{1-Naphthol} & \multicolumn{4}{c}{2-Naphthol} \\
\hline
& $R_{ij1}$ &$q_{ij1}$ & $u_{ij1}^*$ &  $\E[T_{i1}]$ & $R_{ij2}$ &$q_{ij2}$ & $u_{ij2}^*$ &  $\E[T_{i2}]$\\
 NS&    6.39   &   353 & 0 & 1274 &8.63 & 1350 & 1 &1260 \\
 S&	   8.54 & 1366  & 1 & & 7.07 & 363& 0&\multicolumn{1}{c|}{}\\
\cline{2-9}
&\multicolumn{8}{c|}{Minimax}\\

&&&\multicolumn{2}{c}{$\bu^*$} & $\E[T_{i1}]$&  $\E[T_{i2}]$&\multicolumn{2}{c|}{}\\
&&&\multicolumn{2}{r}{$[0.953, 0.391]$} & $571$&  $1137$&\multicolumn{2}{c|}{}\\
\hline

\end{tabular}
\end{center}
\end{table}

To make concrete the factors allowing for the gains discussed in this work, Table \ref{tab:pair} show the values and aligned ranks for $\log_e$ urinary concentrations of 1-naphthol and 2-naphthol for two individuals, one smoker and one nonsmoker, who were matched as a pair by the full match described in Appendix A. Both individuals are Hispanic males aged over 50, are similar in terms of height and weight, and are both exposed to PAHs occupationally, yet the smoker (labeled $S$) has higher levels of 1-naphthol and lower levels of 2-naphthol.

 The tests of both 1-naphthol and 2-naphthol had observed test statistics that were larger than their expectations under Fisher's sharp null with $\Gamma=1$. Hence, the individual  sensitivity analyses will choose the binary vector of $\mathbf{u}^*_k$ such that the individual with the larger observed response is given the value 1, thus having the higher probability of smoking. For 1-naphthol this is the smoker, but for 2-naphthol this is the nonsmoker, as is shown in Table \ref{tab:pair}. Although we do not know the value of this unmeasured confounder, we do know that logically, the unmeasured confounder cannot simultaneously increase the odds that individual 1 smokes relative to individual 2 and the odds that individual 2 smokes relative to individual 1. Simply combining these two sensitivity analyses would ignore the contradictory values of $\mathbf{u}^*_k$. Table \ref{tab:pair} also gives the expectation of the test statistic for the individual outcomes assessed separately at $\Gamma=10$, a value of $\Gamma$ for which the minimax procedure rejects the overall null, but using Holm-Bonferroni to combine sensitivity analyses fails to reject. Conducting sensitivity analyses separately and allowing for an illogical effect of the unmeasured confounder, the worst-case expectations for the contribution from this matched set to the test statistics' expectations are 1274 and 1260 for 1- and 2-naphthol. 

 Recognizing that the unmeasured confounders must have the same impact on odds of treatment for individuals in a matched set yields markedly different results for the overall sensitivity analysis in this pair, as is demonstrated in the section  labeled ``Minimax'' in Table \ref{tab:pair}. First, we note that the values of the unmeasured confounder for both individuals are fractional, an occurrence which is provably impossible when conducting sensitivity analyses for any given outcome \citep{ros90}. This makes the probabilities of assignment to treatment and control much less extreme than they possibly could have been: conditional on one of the two individuals receiving the treatment, the smoker is given a probability of $\exp\{\log(10)0.391\}/(\exp\{\log(10)0.391 \} +\exp\{\log(10).953 \}\}) = 0.22$ of being a smoker, while at $\Gamma=10$ this probability could have been as low as $1/(1+10)$ and as high as $10/(1+10)$. In minimizing the maximal deviate, the optimization problem determined that a compromise should be made between the two conflicting desires of the individual level sensitivity analysis, but that it should favor making 2-naphthol more significant. Hence, we see that the contribution to the overall expectation of the two test statistics is larger than what it would have been at no unmeasured confounding for 2-naphthol (1137 vs 856.5), but is actually smaller for 1-naphthol (571 vs 859.5).


\subsection{Sensitivity of Overall and Outcome Specific Effects}\label{sec:data}
As was stated in Section \ref{sec:example}, the conclusions of either of the individual level tests on 1- and 2-naphthol were both overturned at $\Gamma= 7.78$ when using Holm-Bonferroni. This is also the maximal level of $\Gamma$ at which we can claim overall significance of at least one of these metabolites. The minimax procedure for testing the overall null hypothesis was able to claim robustness of this same finding up until $\Gamma= 10.22$, representing a substantial increase in robustness. In this application the overall null is of interest, as both naphthalene metabolites are indicators of naphthalene exposure. Hence, rejecting the overall null implies that we can suggest that at least one of our indicators of naphthalene exposure is significantly elevated for smokers relative to nonsmokers, even if we are not able to identify a particular metabolite that is significant at that level of unmeasured confounding.

 To exploit the potential gains in power for individual tests of 1-naphthol and 2-naphthol, we use a closed testing procedure. In our example, doing so means that if we reject the null $H_1\wedge H_2$ at level $0.05$ through our minimax procedure we can then test the individual hypotheses $H_1$ and $H_2$ at level $0.05$ (rather than $0.025$) and still maintain the proper familywise error rate. Since our test of the overall null rejects until $\Gamma = 10.22$, the closed testing procedure allows us to perform individual tests up to that level of unmeasured confounding. The individual tests of 1- and 2-naphthol \textit{without} a Bonferroni correction (i.e., tested at $\alpha = 0.05$) were not overturned until a $\Gamma$ of $7.83$ and $8.20$ respectively. As our minimax  procedure rejects the overall null $H_1\wedge H_2$ for all $\Gamma$ between $7.78$ and $8.20$, we can declare improved robustness of the individual level tests. That is, we can reject the null of no effect for 1- and 2-naphthol at all levels of $\Gamma$ up to $\Gamma=$ $7.83$ and $8.20$, rather than $\Gamma = 7.78$.

 \section{Discussion}

In a randomized clinical trial, counfounders not accounted for in the trial's design are, on average, balanced through randomized assignment of the intervention. As such, there is less of a concern that the observed results are driven by a causal mechanism other than the one under investigation. In observational studies, there is no such guarantee of balance on the unmeasured confounders between the two groups under comparison. When testing for a causal effect on multiple outcome variables,  concerns about a loss of power by controlling the familywise error rate both under the assumption of no unmeasured confounding and within the sensitivity analysis may arise. We have demonstrated through this work that when dealing with multiple comparisons in a sensitivity analysis, the loss in power from controlling the familywise error rate can be attenuated.

As mentioned in Section \ref{sec:closed}, our method can be used in conjunction with sequential rejection procedures which proceed by rejecting intersection null hypotheses on a sequence of subsets of outcomes, $\{\mathcal{K}_\ell\}$. For certain types of null hypotheses, such as those for the value of an additive treatment effect with one sided alternatives, our method could also be used while employing the partitioning principle of familywise error control \citep{fin02}. One deficiency of our method is that it does not account for correlation between test statistics, which can greatly improve power in the presence of dependence \citep{wes93, rom05}. While the simulation studies of Section \ref{sec:simulation} reveal marked improvements when test statistics are independent, these gains are far more modest when the test statistics are correlated and further improvements are desired. Deriving methods for sensitivity analyses which exploit correlation between test statistics remains a topic of ongoing research. Another limitation is that our method can only be used for sensitivity analyses after matching, as the structure of matched sets returned by matching algorithms allows for a straightforward relationship between the assignment probabilities and the variances of our test statistics. In unmatched or stratified analyses, while the logical inconsistencies noted herein are still present, optimizing over the unknown assignment probabilities can no longer be expressed as a quadratically constrained linear program.

In our motivating example, we argue that if smoking causes increased naphthalene exposure, it would elevate levels of both 1- and 2-naphthol in the body. Though related, these metabolites are not affected equally by measured and unmeasured confounding variables: for example, there are certain genetic variants that are only believed to affect the prevalence of particular naphthalene metabolites \citep{yan99}. When focusing on a single outcome variable, the worst-case confounder is allowed to optimally align itself with the responses in each matched set through selecting the worst-case allocation of treatment assignment probabilities. If we are instead trying to disprove the overall truth of null hypotheses on multiple outcomes, the worst-case confounder likely cannot affect the treatment assignment probabilities in a way that simultaneously yields the worst-case inference for all outcomes. Exploiting this fact not only lends higher power to a sensitivity analysis for the overall null across all outcomes, but also increases power for testing hypotheses on individual outcomes through the use of certain sequential rejection procedures.

\begin{center}
{\large\bf SUPPLEMENTARY MATERIAL}
\end{center}
\begin{description}
\item[Appendices] Appendix A describes the details of the matching performed in our motivating example on the impact of smoking on naphthalene levels. Appendix B discusses how our procedure can be extended to test hypotheses with one-sided alternatives. Appendix C contains a simulation study demonstrating that our proposed procedure strongly controls the familywise error rate (.pdf file).
\item[\texttt{R}-script] \texttt{multiCompareFunctions.R} provides functions for conducting a sensitivity analysis for any intersection null through the solution of a quadratically constrained linear program.
\item[\texttt{R}-script] \texttt{reproduceScript.R} provides code for reproducing the results of this paper.
\item[Data set] \texttt{naphthalene.csv} provides the data used in this paper.
\end{description}
\newpage

\begin{center}
{\Large APPENDIX}

\end{center}
\appendix
\section{Additional Details for the Smoking and Naphthalene Example}
\doublespacing
 Following \citet{wei05} and \citet{suw09}, individuals were classified as active smokers if they stated that they smoke ``every day'' or ``some days''  in response to the question ``Do you now smoke cigarettes?,'' or if their serum cotinine (a metabolite of nicotine) levels were above 0.05 ng/mL. Using this definition, there were 453 smokers and 1253 nonsmokers.  The nonsmokers include former smokers and never smokers, as urinary naphthol is an indicator of recent naphthalene exposure. 
 
\begin{figure}[h]
\begin{center}
\includegraphics[scale = .6]{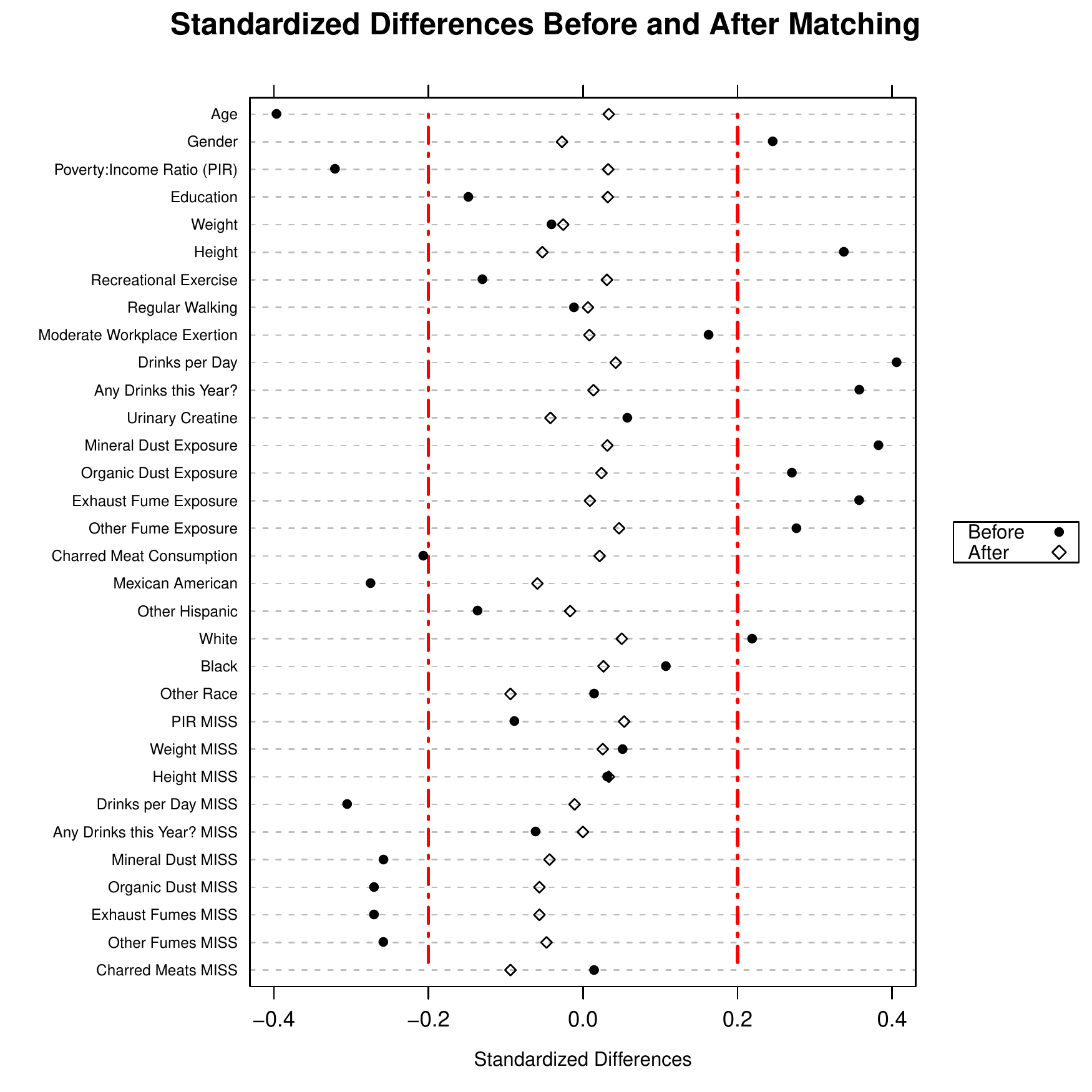}
\caption{\small{Covariate Imbalances Before and After Matching. The dotplot (a Love plot) shows the absolute standardized differences without matching, and after conducting a matching with a variable number of controls. The vertical dotted line corresponds to a standardized difference threshold of 0.2, which is often regarded as the maximal allowable absolute standardized difference \citep[for example,][]{sil01}}. The largest absolute standardized difference after matching was 0.094.}
\label{fig:balance}
\end{center}
\end{figure}

 We used full matching to control for observed covariates in this study \citep{ros91, han04}. In this match, we allowed for strata of maximal size 10, meaning that a matched set could have, at most, either 1 current smoker and 9 nonsmokers; or 1 nonsmoker and 9 current smokers. We identified 22 pre-treatment covariates deemed predictive of smoking and naphthalene levels based on those used in \citet{suw09}; these covariates are listed in Figure \ref{fig:balance}. Ten covariates contained missing values, with a maximal percentage of values missing of 10\%.  To account for this, we included 10 missingness indicators as additional covariates upon which to match. As discussed in \citet{ros84} and \citet[][Section 9.4]{designofobs}, this facilitates balancing the observed covariates and the pattern of missingness. Rank-based Mahalanobis distance with a propensity score caliper of 0.08 was used, and propensity scores were estimated using logistic regression \citep[][Section 8.3]{designofobs}.   Figure \ref{fig:balance} shows the standardized differences before and after matching for observed confounders and demonstrates that before matching there were substantial imbalances between smokers and nonsmokers with respect to many important variables. It also shows that matching was able to effectively create a well-balanced comparison between smokers and nonsmokers on the basis of these variables. Details for calculating standardized differences before and after full matching can be found in \citet{stu08} and \citet[][Section 9.1]{designofobs}.

\section{A Simple Extension To One-Sided Testing}
\doublespacing
By taking the square of the deviate in our original formulation, we lose the deviate's sign. While this does not make a difference for two-sided testing, it does make a difference when the test is one-sided. For example, if we stipulated a one-sided, greater than alternative  but observed a test statistic markedly \textit{smaller} than its expectation under the null we should fail to reject that null, a fact which is lost when taking the square. To account for this, we introduce a penalty into the constraints corresponding to one-sided hypotheses that only allow for a rejection to be registered if the expectation of the test statistic yielded through the sensitivity analysis maintains the proper relationship with the observed test statistic given the nature of the alternative. Let $b_k$ be a binary variable for the $k^{th}$ outcome, and let $M$ be a sufficiently large constant. 

Redefine $\zeta_k(\brho)$ so that

\begin{align*} \zeta_k(\brho) &=  \begin{cases}(t_k - \E[t_k(\bZ, \bF_k);\brho])^2 -  \chi^2_{1,1-\alpha/K}\Var(t_k(\bZ, \bF_k);\brho) & \text{if two-sided alternative}\\
(t_k - \E[t_k(\bZ, \bF_k);\brho])^2 -  \chi^2_{1,1-2\alpha/K}\Var(t_k(\bZ, \bF_k);\brho) & \text{if one-sided alternative}\end{cases} \end{align*}

We then modify our optimization problem as follows:

\begin{align*}
\underset{{y,\varrho_{ij},s_i, b_k} }{\text{minimize}} &\;\; y\\
\text{subject to}\;\; & y \geq \zeta_k(\brho) - Mb_k\;\; \forall k\\
& \;\; \sum_{j=1}^{n_i}\varrho_{ij} = 1 \;\;\forall i \nonumber\\
&s_i\leq \varrho_{ij} \leq \Gamma s_i\;\; \forall i,j\\
&\varrho_{ij} \geq 0 \;\; \forall i,j\\
&b_k \in \{0,1\}\;\; \forall k\\
&b_k = 0 \;\;\text{if }H_k\text{ two-sided}\\
-M(1-b_k) \leq t_k - \brho^T\bq_k &\leq Mb_k\;\;\text{if }H_k\text{ one-sided }, <\\
-Mb_k \leq t_k - \brho^T\bq_k &\leq M(1-b_k)\;\;\text{if }H_k\text{ one-sided }, >\\
\end{align*}

The value $M b_k$ added to the $k$ constraints on the auxiliary variable $y$, in conjunction with the constraints on the value of the test statistic's numerator, impose a heavy negative penalty if the relationship between the test statistic and its mean under a given allocation of unmeasured confounders do not adhere to the required direction imposed by the alternative. This makes it such that we will never reject a null at a given $\Gamma$ because a given one-sided test was highly \textit{insignificant}, which without such a penalty would be construed as being highly significant.

\section{Simulation of Type I Error Control}
In this simulation study, we demonstrate that, in the presence of true intersection null hypotheses,  our procedure strongly controls the familywise error rate at level $\alpha = 0.05$.  In each of 6 simulation settings, we simulate 10,000 data sets under no unmeasured confounding with $I=250$ pairs for three outcome variables of interest and using Huber's M-statistic, as described in Section 5.1 of the manuscript. For each of the 2 combinations of treatment effects and covariances, closed testing is used, with our minimax procedure being used for each intersection null. Tests are run at $\Gamma =1, 1.05$, and $1.1$. The values for the treatment effect vector and the covariances were as follows:

\begin{enumerate}
\item $\btau = [0,0,0.3]$
\item $\Sigma^{(1)} = \text{Diag}(1)$; $\Sigma^{(2)}_{ij} = 1$ if $i=j$, $\Sigma^{(2)}_{ij} = 0.5$ otherwise.
 \end{enumerate}

  \begin{table}
  \begin{center}
\caption{ \label{tab:individ}Rejection probability for testing true and false nulls through closed testing. Desired strong familywise error control at 0.05.}
\begin{tabular}{c c | c c c | c c}
\hline
&\multicolumn{1}{c}{}&\multicolumn{3}{c}{True Nulls} & \multicolumn{2}{c}{False Nulls}\\
\multicolumn{1}{c}{Gamma} & \multicolumn{1}{c}{Moments} & $H_1$ & $H_2$ & \multicolumn{1}{c}{$H_1\wedge H_2$} & $H_3$& \multicolumn{1}{c}{$H_1\wedge H_2 \wedge H_3$} \\
\hline
\multirow{2}{*}{$\Gamma = 1 $}
&	$\btau,	\Sigma^{(1)}$&0.0260
&
0.0266
&
0.0506
&
0.9884
&
0.9886
\\

&	$\btau,	\Sigma^{(2)}$&
0.0267
&
0.0268
&
0.0462
&
0.9881
&
0.9893

\\

\hline																		
\multirow{2}{*}{$\Gamma = 1.05 $}
&	$\btau,	\Sigma^{(1)}$&
0.0102
&
0.0089
&
0.0189
&
0.9748
&
0.9749

\\

&	$\btau,	\Sigma^{(2)}$&
0.0096
&
0.0122

&
0.0197
&

0.9732
&
0.9750

\\

\hline																		
\multirow{2}{*}{$\Gamma = 1.10 $}
&	$\btau,	\Sigma^{(1)}$&
0.0035
&
0.0043

&
0.0078
&
0.9462
&
0.9463
\\

&	$\btau,	\Sigma^{(2)}$&
0.0053
&
0.0032
&
0.0081
&
0.9441
&
0.9462
\\

\hline
 \end{tabular}
  \end{center}
 \end{table}

We test Fisher's sharp null on each outcome. In each iteration, we record whether or not the true null hypotheses $H_1$, $H_2$, and $H_1\wedge H_2$ are rejected. We also record whether or not the false nulls $H_3$ and $H_1\wedge H_2\wedge H_3$ are rejected.  Table \ref{tab:individ} shows the results of this simulation study. As can be seen, our procedure strongly controls the type I error rate for all values of $\Gamma$ tested. The rate of rejection for $H_1\wedge H_2$ reveals that our procedure is conservative when the test statistics are dependent, while coming very close to attaining the actually desired level under independence. As $\Gamma$ increases the Type I error rate decreases for all true nulls, as many spurious rejections assuming no unmeasured confounding can be explained by moderate departures from pure randomization.

\bibliographystyle{apalike}
\bibliography{bibliography}
\end{document}